\documentclass[sigconf]{acmart}
\usepackage{marginnote}
\usepackage{booktabs}
\usepackage{marvosym}
\usepackage{enumitem}
\usepackage[ruled,linesnumbered]{algorithm2e}
\usepackage{amsmath}
\usepackage{braket}
\usepackage{bm}
\usepackage[title]{appendix}
\usepackage{multirow}
\usepackage{color}
\AtBeginDocument{%
  \providecommand\BibTeX{{%
    \normalfont B\kern-0.5em{\scshape i\kern-0.25em b}\kern-0.8em\TeX}}}

\setcopyright{acmcopyright}
\copyrightyear{2022}
\acmYear{2022}
\acmDOI{10.1145/3534678.3539130}

\acmConference[KDD'22]{Proceedings of the 28th ACM SIGKDD Conference on Knowledge Discovery and Data Mining}{August 14--18, 2022}{Washington, DC, USA.}
\acmBooktitle{Proceedings of the 28th ACM SIGKDD Conference on Knowledge Discovery and Data Mining (KDD '22), August 14--18, 2022, Washington, DC, USA}
\acmPrice{15.00}
\acmISBN{978-1-4503-9385-0/22/08}

\begin{document}

\title{Feature-aware Diversified Re-ranking with Disentangled Representations for Relevant Recommendation}

\settopmatter{authorsperrow=4}

\author{Zihan Lin$^{\dagger}$}
\email{zhlin@ruc.edu.cn}
\affiliation{
  \institution{School of Information, Renmin University of China}
  \city{}
  \country{}
}
\author{Hui Wang$^{\dagger}$}
\email{hui.wang@ruc.edu.cn}
\affiliation{
  \institution{School of Information, Renmin University of China}
  \city{}
  \country{}
}

\author{Jingshu Mao}
\email{maojingshu@kuaishou.com}
\affiliation{
  \institution{Kuaishou Inc.}
  \city{Beijing}
  \country{China}
}
\author{Wayne Xin Zhao$^{\clubsuit}$ $^{\spadesuit}$\textsuperscript{\Letter}}
\email{batmanfly@gmail.com}
\affiliation{
  \institution{Gaoling School of Artificial Intelligence, Renmin University of China}
  \city{}
  \country{}
}
\author{Cheng Wang}
\email{wangcheng03@kuaishou.com}
\affiliation{
  \institution{Kuaishou Inc.}
  \city{Beijing}
  \country{China}
}
\author{Peng Jiang}
\email{jiangpeng@kuaishou.com}
\affiliation{
  \institution{Kuaishou Inc.}
  \city{Beijing}
  \country{China}
}
\author{Ji-Rong Wen$^{\clubsuit}$ $^{\spadesuit}$}
\email{jrwen@ruc.edu.cn}
\affiliation{
  \institution{Gaoling School of Artificial Intelligence, Renmin University of China}
  \city{}
  \country{}
}
\thanks{$\dagger$ Equal contribution. Work done during internship at KuaiShou.}
\thanks{$\spadesuit$ Beijing Key Laboratory of Big Data Management and Analysis Methods}
\thanks{$\clubsuit$ Beijing Academy of Artificial Intelligence, Beijing, 100084, China}
\thanks{\textsuperscript{\Letter} Corresponding author.}

\renewcommand{\authors}{Zihan Lin, Hui Wang, Wayne Xin Zhao, Cheng Wang, Peng Jiang, Ji-Rong Wen}
\renewcommand{\shortauthors}{Lin and Wang, et al.}

\newcommand{\todo}[1]{\textcolor{blue}{#1}}
\newcommand{\ie}{\emph{i.e.,}\xspace}
\newcommand{\eg}{\emph{e.g.,}\xspace}
\newcommand{\aka}{\emph{a.k.a.,}\xspace}
\newcommand{\etal}{\emph{et al.}\xspace}
\newcommand{\paratitle}[1]{\vspace{1.5ex}\noindent\textbf{#1}}
\newcommand{\wrt}{w.r.t.\xspace}
\newcommand{\ignore}[1]{}
\newcommand{\ourmodel}{FDSB~}
\begin{abstract}
Relevant recommendation is a special recommendation scenario which provides relevant items when users express interests on one target item~(\eg click, like and purchase). Besides considering the relevance between recommendations and trigger item, the recommendations should also be diversified to avoid information cocoons. However, existing diversified recommendation methods mainly focus on item-level diversity which is insufficient when the recommended items are all relevant to the target item. Moreover, redundant or noisy item features might affect the performance of simple feature-aware recommendation approaches. 

Faced with these issues, we propose a \textbf{F}eature \textbf{D}isentanglement \textbf{S}elf-\textbf{B}alancing Re-ranking framework~(\textbf{FDSB}) to capture feature-aware diversity. The framework consists of two major modules, namely disentangled attention encoder~(DAE) and self-balanced multi-aspect ranker. In DAE, we use multi-head attention to learn disentangled aspects from rich item features. In the ranker, we develop an aspect-specific ranking mechanism that is able to adaptively balance the relevance and diversity for each aspect. 
In experiments, we conduct offline evaluation on the collected dataset and deploy \ourmodel on KuaiShou app for online $A$/$B$ test on the function of relevant recommendation. The significant improvements on both recommendation quality and user experience  verify the effectiveness of our approach.
\end{abstract}



\maketitle

\section{Introduction}

Nowadays, recommender systems are playing an increasingly important role in delivering suitable content to users and boosting the exposure of high-quality resources~\cite{ricci2011introduction}. From classic collaborative filtering~\cite{su2009survey} to recent neural recommenders~\cite{gupta2020deeprecsys}, the underlying recommendation algorithms~\cite{RecBole} have been evolving during the past two decades, for better satisfying  user preferences.
Besides the algorithmic aspect, various auxiliary functionalities have been introduced in recommender systems to further improve user experiences. 
Among these new functionalities, \emph{relevant recommendation} has been widely deployed in e-commerce platforms and video platforms~\cite{xie2021real}, both recommend relevant items~(products or videos) of a specific item~(called \emph{trigger item}).
For example, Youtube uses ``Up next'' section to show a suggestion of what to watch next when you are watching a video. Amazon would present relevant products~(``\emph{Products related to this item}'') on product description page. With relevant recommendations, Amazon is able to keep consumers engaged and offer products of interest to them that they might not even think about before. 

\begin{figure}[t]
    \centering
    \includegraphics[width=0.45\textwidth]{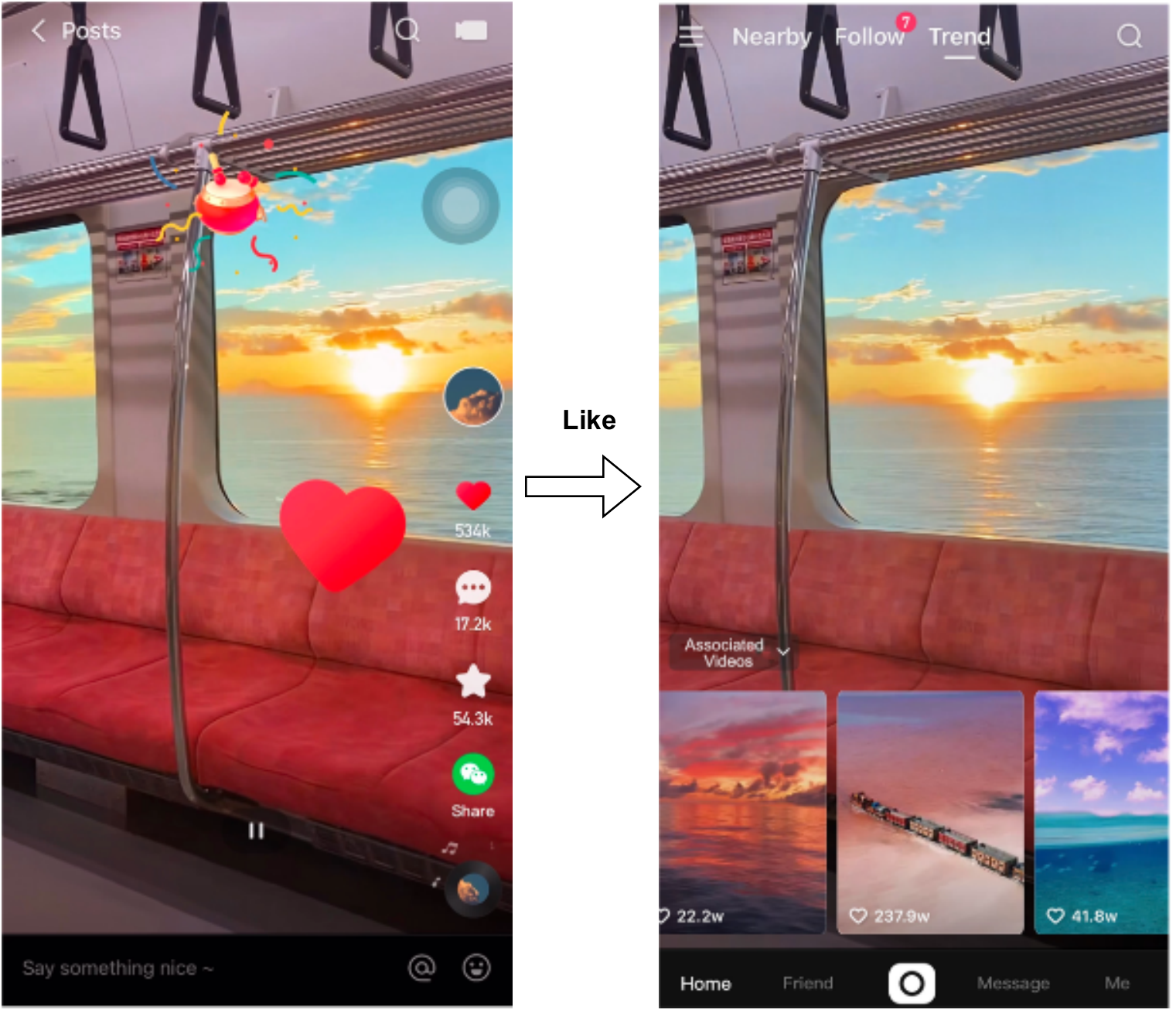}
    \caption{A snapshot of the relevant recommendations in KuaiShou. Left: The video which contains three feature tags~(\eg \emph{sunset}, \emph{train} and \emph{sea}~) is liked by a user. Right: A small number of relevant videos are displayed below.}
    \label{fig:intro}
\end{figure}

In this paper, we study the task of \emph{relevant recommendation} in a special scenario of micro-video recommendation on \emph{KuaiShou}\footnote{\url{https://www.kuaishou.com/}}, which is a popular Chinese micro-video app (similar to \emph{TikTok}). 
Compared with other types of apps, it is easier to activate a consecutive watching of videos, since it takes less time to watch micro-videos than ordinary videos. 
We present an illustrative snapshot of this function on Kuaishou app in Figure \ref{fig:intro}.
When a user clicks on the `\emph{like}' button for a micro-video (the left part of Figure~\ref{fig:intro}), the system will present a small number of micro-videos for recommendation (the right part of Figure~\ref{fig:intro}). Furthermore, these recommended micro-videos can be directly clicked for watching. Generally speaking, we refer to the micro-video being liked \emph{trigger item} and the recommended micro-videos as \emph{relevant items}.

The main purpose of relevant recommendation is to recommend both \emph{relevant} and \emph{diverse} items. For relevance,  the videos recommended should be relevant to the trigger item, since the user expresses positive preference on it.  For diversity, the recommendations should be diverse to include new features for avoiding information cocoons, and explore more interests of users.
In fact, such a task setting is similar to \emph{diversified recommendation}~\cite{wu2019recent}. A significant difference is that relevant recommendation is conducted based on both the trigger item and user preference, while diversified recommendation only considers the user preference.  To incorporate the trigger item for diversified recommendation, we need to explicitly model fine-grained item characteristics in order to better characterize item-to-item relations. For example, in Figure \ref{fig:intro}, the liked micro-video is attached with a set of semantic labels \{~\emph{sunset}, \emph{train} and \emph{sea}~\}, and it is meaningful to conduct  the relevant recommendations according to these fine-grained semantic labels (\ie features). 
However, existing diversified recommendation methods mainly focus on item-level diversity~\cite{huang2021sliding}, which is not able to capture feature-level diversity effectively.  In addition, these side information can be rather redundant or noisy, and a direct feature-level recommendation approach (\eg simple finding and merging relevant items for each feature) may not work well in practice~\cite{cheng2022feature}.   

Considering the above issues, we design a general re-ranking framework, \textbf{F}eature \textbf{D}isentanglement \textbf{S}elf-\textbf{B}alanced Re-ranking (\textbf{FDSB}), to capture \emph{feature-aware diversity} for  relevant recommendation.
Our framework is developed based on a multi-factor re-ranking paradigm, 
consisting of two major modules, namely disentangled attention encoder and self-balanced multi-aspect re-ranker. 
For the disentangled attention encoder, we learn disentangled dimensions from rich item features, called \emph{aspects}, based on multi-head attention. The learned aspects provide a more compact representation to extract item features, which will be used in subsequent re-ranking. For the self-balanced multi-aspect re-ranker, we develop an aspect-specific re-ranking mechanism that is able to adapt balance relevance and diversity.
We set aspect-specific coefficients for each dimension to tune the importance of both targets, and associate these coefficients of relevance and diversity for balancing the two factors.

To verify the effectiveness of proposed framework, we conduct both offline evaluation experiments on a large user logged dataset and online A/B testing experiments on the Kuaishou app. Experimental results show that our proposed framework outperforms the comparison baselines in recommendation quality and is more suitable to promote both relevance and diversity for relevant recommendation.
Our contributions can be summarized as follows:

\begin{itemize}[leftmargin=*]
    \item We formulate the task of relevant recommendation in a multi-factor ranking paradigm, which jointly considers the user preference, relevance and diversity for re-ranking. 
    
     \item  We propose a feature disentanglement self-balanced re-ranking framework for this task, which learns more compact aspects from rich item features and conducts multi-aspect adaptive ranking to balance different factors. 
     
     \item  We demonstrate the effectiveness  of proposed framework through both offline experiments and online $A/B$ test. 
\end{itemize}
\section{Methodology}
In this section, we present the proposed \ourmodel framework for relevant recommendation. We first define the notations and formulate the task in Section~\ref{sec:task}. Then we introduce the disentangled attention encoder in Section~\ref{sec:dae}, which encodes multiple features into different representations. After that, a feature-aware diversified recommendation algorithm is proposed in Section~\ref{sec:sm3r}. Finally, we  discuss the proposed framework in Section~\ref{sec:discussion}. 

\subsection{Task Formulation}\label{sec:task}

The task of \emph{relevant recommendation} aims to return a set of suitable items to a target user given a trigger item. Since such a task is seldom considered by previous work~\cite{xie2021real}, we first formulate the task and present an approximate greedy approach. 

\subsubsection{A Multi-Factor Ranking Formulation}
We formulate the task of relevant recommendation as a multi-factor ranking problem, where the preference score of users, the relevance to the trigger item and the diversity of the returned recommended items are considered for a balanced ranking. 
Formally, let $\mathcal{I}$ denote the full item candidate set for recommendation, $i_t$ denote the trigger item and $u$ denote a target user. This task can be described as an conditional subset selection problem~\cite{daszykowski2002representative} from the item set $\mathcal{I}$ as follows:
\begin{equation}
    \mathcal{R} = \arg \max_{\mathcal{I}{'} \subset \mathcal{I}}  \underbrace{\text{Pref}(u, \mathcal{I}{'})}_{Preference} + \lambda \cdot \underbrace{\text{Rel}(\mathcal{I}{'}, i_t)}_{Relevance} + \underbrace{\gamma \cdot \text{Div}(\mathcal{I}{'} | i_t)}_{Diversity}, \label{eq:overall}
\end{equation}
where $\mathcal{R}$ is the returned set of relevant recommended items to user $u$ given the trigger item $i_t$ and $\mathcal{I}{'}$ denotes the selected items. Here, $\text{Pref}(\cdot)$ represents the preference score of user $u$ to returned items,  $\text{Rel}(\cdot)$ measures the relevance between trigger item $i_t$ and $\mathcal{I}{'}$, and $\text{Div}(\cdot)$ calculates the diversity of returned items in term of trigger item. We set two hyper-parameters $\lambda$ and $\gamma$ to balance the weights of the three factors.

\subsubsection{Feature-aware Relevance and Diversity }
In the above formulation, we can utilize existing recommendation methods to compute the preference term $\text{Perf}(u, \mathcal{I}{'})$. While, the major difficulty lies in how to derive the latter two terms, \ie \emph{relevance} and \emph{diversity}, which is the focus of this paper. To model these two terms, we consider a feature-aware way, where we assume the side information of items (\eg brand, color and avatar) is available for recommendation, denoted as a feature set  $\mathcal{F}_i$ for item $i$.
Furthermore, we assume that a pre-trained item embedding, $\bm{e}_i$, can also be used for re-ranking.
In this setting, the terms of relevance and diversity can be defined as:
\begin{eqnarray}
    \text{Rel}(\mathcal{I}', i_t) &=& \sum_{i' \in \mathcal{I}'} \text{sim}^I(\bm{e}_{i'}, \bm{e}_{i_t}) + \text{sim}^F(\mathcal{F}_{i' }, \mathcal{F}_{i_t}), \label{eq:def_rel}\\
    \text{Div}(\mathcal{I}{'}| i_t) &=& \text{Div}^I(\{\bm{e}_{i{'} }|i{'} \in \mathcal{I}{'}\}) + \text{Div}^F(\{\mathcal{F}_{i{'} }\cap \mathcal{F}_{i_t}|i{'} \in \mathcal{I}{'}\}),\nonumber 
\end{eqnarray}
where $i{'} \in \mathcal{I}{'}$ is a candidate item to be recommended, and the two factors are modeled at both item level (with the superscript of $I$) and feature level (with the superscript of $F$).  For relevance, it is calculated according to the item embedding similarity ($\text{sim}^I(\cdot)$) and the feature similarly ($\text{sim}^F(\cdot) $). Similarly, the diversity is computed according to embedding diversity ($\text{Div}^I(\cdot)$) and feature diversity ($\text{Div}^F(\cdot)$). 
Most of existing works mainly focus on item-level relevance or diversity~\cite{xie2021real,huang2021sliding}. Here, we consider more fine-grained characterization of the two factors at the feature level. 

\subsubsection{Greedy  Approximation}

Above, we have formulated the diversified re-ranking problem of relevant recommendation as a combinatorial optimization problem based on item embeddings and features in Eq.~\eqref{eq:overall} and Eq.~\eqref{eq:def_rel}. However, this problem has been proved to be NP-hard~\cite{toth2000optimization} for exact solutions. So, we turn to a greedy selection approach~\cite{nemhauser1978analysis} that decomposes the set based objective into a single-item objective according to 
the marginal gain as follow:
\begin{equation}
    i_l = \arg\max_{i \in \mathcal{I}\backslash \mathcal{R}_{1:l-1}} \text{Pref}(u, i) + \lambda \cdot \text{Rel}(i, i_t) + \gamma\cdot \text{Div}(\mathcal{R}_{1:l-1}\cup i| i_t), \label{eq:marginal_overall}
\end{equation}
where $l$ is the current selection step and we make the selection based on the results of previous $l-1$ steps. 
Through such a greedy approximation, we can easily compute the second term of relevance score, and the key is how to model the third term of diversity, considering both item-level and feature-level diversity. This is a major difference compared with previous works~\cite{carbonell1998use,chen2018fast,huang2021sliding} which mainly focus on item-level diversity. 

To address this difficulty, we design a novel relevant recommendation framework, consisting of two major modules, namely \emph{disentangled attention encoder} that disentangles the features into different aspects and \emph{self-balanced multi-aspect re-ranker} that balances the diversity of multiple feature aspects. In what follows, we describe the two parts in detail. 

\subsection{Disentangled Attention Encoder}\label{sec:dae}

\begin{figure}[t]
    \centering
    \includegraphics[width=0.38\textwidth]{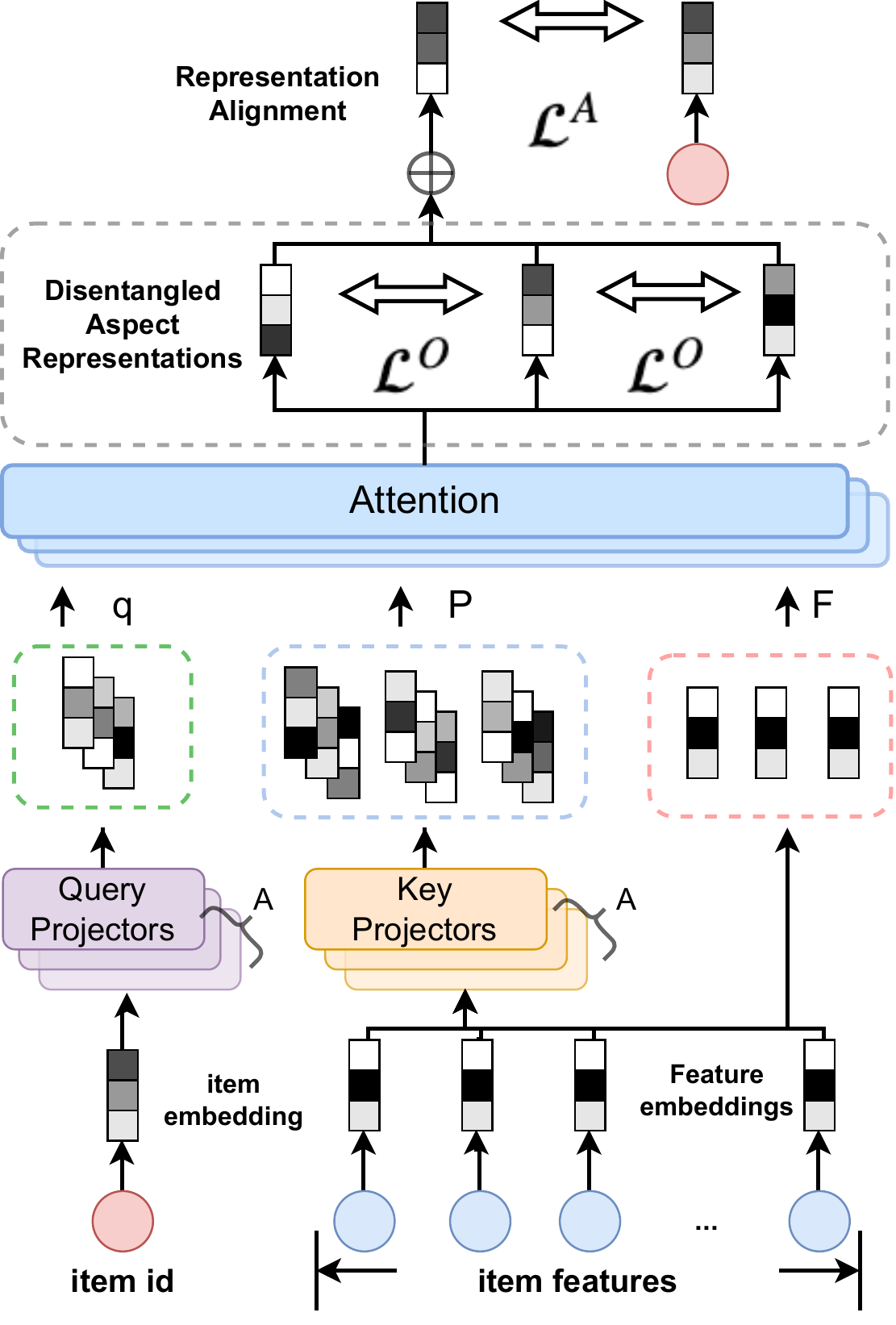}
    \caption{Architecture of our proposed disentangled attention encoder~(DAE), which takes pre-trained item embedding and feature set as input and produces multiple disentangled representations (\ie aspects).}
    \label{fig:decomp}
\end{figure}

In recommender systems, the item resources are usually attached with rich feature information, which correspond to different latent aspects describing the item characteristics from different perspectives. However, these features are usually correlating or redundant. 
Considering this issue, we propose a disentangled attention encoder~(DAE) to disentangle these features into fine-grained feature-aware representations corresponding to different aspects. This module is the base to model the terms of relevance and diversity in our framework. The overall architecture of DAE is depicted in Figure~\ref{fig:decomp}.

\subsubsection{Attention-based Feature Disentanglement}

For an item $i$, an embedding $\bm{e}_i \in \mathbb{R}^{d}$ (either pre-trained or to be learned) is associated. For simplicity, we omit the index of item $i$ in the following notations. Besides, it corresponds to a feature set of side information, denoted as $\mathcal{F} =\{f^1,f^2,\ldots,f^F\}$. 
Following the idea of disentangled representation learning~\cite{tran2017disentangled}, we aim to learn multi-aspect disentangled representations over the feature set, denoted as $\bm{V} = \{\bm{v}^1,\bm{v}^2,\ldots,\bm{v}^A\}$, where $A$ is the number of disentangled dimensions (\ie aspects). 

To learn the disentangled representations, instead of following existing approaches~\cite{ma2019learning,mu2021knowledge}, we adopt a simpler approach suited to our setting based on the multi-head attention~\cite{vaswani2017attention} to conduct disentangled representation learning.
The input query vectors and key vectors are projected while the value vectors remain fixed across different heads. Specifically, input features are firstly mapped to learnable vectors through an embedding layer:
\begin{equation}
    \bm{F} = \text{Emb}(\mathcal{F}),
\end{equation}
where $\bm{F} \in \mathbb{R}^{F \times d}$ is the initial embedding matrix of $F$ features. Then, item embeddings and feature embeddings are projected $A$ times in parallel corresponding to $A$ aspect representations where the $a$-th projected vector is produced by multi-layer perceptrons~(MLP):
\begin{equation}
\begin{aligned}
    \bm{q}^a &= \text{MLP}_{Q}^a(\bm{e}_i),\\
    \bm{P}^a &= \text{MLP}_{K}^a(\bm{F}).
\end{aligned}
\end{equation}
After that, we use the original feature embeddings as value and projected vectors as query and key to calculate attention scores. The $a$-th aspect representation is computed by scaled attention:
\begin{equation}
    \bm{v}^{a} = \text{Attention}(\bm{q}^a, \bm{P}^a, \bm{F}). \label{eq:attention}
\end{equation}
The weights of each feature in the attention block are calculated by the scaled dot product of projected query vector and key matrix in the form of $\text{Softmax}(\frac{\bm{q}^a \cdot \bm{P}^a}{\sqrt{d}})$.

\subsubsection{Alignment with Item Representations}

Since these aspects capture part of characteristics of items, their representations should be closer to the overall item embedding in order to reflect item-specific characteristics. 
Based on this idea, we design two semantic alignment methods for distilling the information of item embedding into aspect representations:
\begin{eqnarray}
\mathcal L^{A}_{MSE} &=& \frac{1}{2} \parallel \bm{e}_i-\widetilde{\bm{v}} \parallel_2^2, \\
\mathcal L^{A}_{InfoNCE} &=&  -\log \frac{\exp(\bm {e}_i\cdot \widetilde{\bm{v}}/\tau)}{\sum_{i{'} \in \mathcal{I}\backslash i} \exp(\bm {e}_{i{'}} \cdot \widetilde{\bm{v}}/\tau)},
\end{eqnarray}
where $\widetilde{\bm{v}} = \frac{1}{A} \sum_{a=1}^{A} \bm{v}^{a}$ is the average of the aspect representations. The first alignment loss is based on  point-wise Mean Squared Error~(MSE) loss while the second alignment loss conducts the InfoNCE loss~\cite{oord2018representation} by 
treating other items in the same batch as negative examples. Besides the alignment loss, we further require that the disentangled aspects capture different semantic characteristics. Therefore, we introduce the orthogonalization loss $\mathcal L^{O}$ which is conducted in a similar contrastive manner: 
\begin{equation}
    \mathcal L^{O} = \sum_{a=1}^A -\log \frac{\exp(\bm{v}^a \cdot \bm {v}^a/\tau)}{\sum_{a{'}=1}^A \exp(\bm{v}^{a{'}} \cdot \bm{v}^a/\tau)},
\end{equation}
where $\tau$ is a temperature parameter and $A$ is the number of disentangled aspects representations. By optimizing the above two parts jointly, we can efficiently disentangle rich feature information into compact feature-aware representations.

\subsection{Self-balanced Multi-aspect Re-Ranker}\label{sec:sm3r}

A fundamental approach to diversity ranking is the \emph{Maximal Marginal Relevance}~(MMR)~\cite{carbonell1998use}, and a number of studies have been proposed to improve and generalize the MMR-based algorithm~\cite{di2014analysis,ashkan2015optimal}. 
In our setting, we aim to capture the diverse characteristics corresponding to different aspects (\ie the disentangled representations based on item features). For this purpose, we extend the classic MMR framework and design a self-balanced multi-aspect re-ranker that is able to adaptively tune the aspect weight for diversity re-ranking, summarized in Algorithm \ref{alg:sm3r}. 

\subsubsection{Aspect-specific Adaptive Balance}
In order to model relevance and diversity over different aspects, we introduce learnable  coefficient vectors
$\bm{w}^R \in \mathbb{R}^{A}$ and $\bm{w}^D \in \mathbb{R}^{A}$, to adaptively integrate different aspects. 
To guide the update of combination vectors at each selection step, we calculate the accumulated relevance (denoted as $\bm{r} \in \mathbb{R}^{A}$) before each selection. Furthermore, let $r_a$ denote the similarity between the proceeding selections $\mathcal{R}_{1:l-1}$ and the trigger item $i_t$ for the $a$-th aspect. For multi-factor ranking, the major difficulty lies in the balance between relevance and diversity, since the two factors are essentially contradictive: the more relevant to the trigger item, the less diverse the relevant recommendations are, and vice versa. 

Intuitively, we should enhance the weight of relevance when diversity is well captured, and reduce the weight of relevance  when diversity is less well captured. There should be a trade-off between the two factors.  Since we consider the balance at the aspect level, we design a relevance-diversity relevant weighting mechanism that adaptively captures the balance as follows:
\begin{eqnarray}
    \bm{w}^D &=& \text{softmax}(\bm{r}), \label{eq:div_weight}\\
    \bm{w}^R &=& \text{softmax}(\frac{1}{\bm{r}}), \label{eq:rel_weight}
\end{eqnarray}
where $\frac{1}{\bm{r}}$ denotes the vector by performing 
the element-wise reciprocal computation for $\bm{r}$, and
 each entry in $\bm{r}$ is computed by  $r_a = \sum_{i \in \mathcal{R}} \bm{v}_i^a \cdot \bm{v}_{i_t}^a$ where $\bm{v}_i^a$ is the disentangled representation of item $i$ in $a$-th aspect which is calculated by Eq.~\eqref{eq:attention}.
By adaptively updating the coefficients of multiple aspects, the selection process can be conducted in an aspect-specific adaptive way.

\subsubsection{Greedy Selection}
We firstly consider the selection of the first item $i_1$. In this simple case, 
since there are no selected items (\ie $\mathcal{R}_{1:l-1}=\emptyset$), 
it degenerates into the single-item relevance score as:
\begin{equation}
i_1 = \arg\max_{i \in \mathcal{I}}  \text{Pref}(u,i) + \lambda \big(\bm{e}_i\cdot \bm{e}_{i_t} + \sum_{a=1}^A \bm{v}^{a}_i \cdot \bm{v}^{a}_{i_t}\big), \label{eq:one}
\end{equation}
where we compute the relevance score in terms of item embeddings and representations for $A$ disentangled aspects.
When $l>1$, the overall objective of relevance and diversity are determined by the selected items, and the self-balanced coefficients are utilized to control the balance in multiple aspects. The relevance scores of each aspect between candidate item $i$ and trigger item $i_t$ are weighted by the aspect-specific coefficients as:
\begin{equation}
    \text{Rel}(i, i_t) = \bm{e}_i\cdot \bm{e}_{i_t} + \sum_{a=1}^A w^R_a\big(\bm{v}^{a}_i \cdot \bm{v}^{a}_{i_t}\big).
    \label{eq:rel}
\end{equation}
Correspondingly, the diversity scores from multiple aspects can be also weighed  as follows:
\begin{equation}
    \text{Div}(\mathcal{R}_{1:l-1}\cup i| i_t) = 1- \max_{i{'} \in \mathcal{R}_{1:l-1}} \bigg(\bm{e}_{i{'}}\cdot \bm{e}_i+ \sum_{a=1}^A w^D_a  \big(\bm{v}_{i'}^{a}\cdot \bm{v}_i^{a}\big)\bigg).
    \label{eq:div}
\end{equation}
In this diversity function, we subtract the maximal similarity between selected items and candidate item as the diversity score which follows the  MMR approach~\cite{carbonell1998use}. 
While, other choices of diversity functions can be easily adapted, which will be discussed in the following. After defining the relevance and diversity scores, the $l$-th item can be selected according to Eq.~\eqref{eq:marginal_overall}.

\begin{algorithm}
    \caption{The algorithm for self-balanced multi-aspect re-ranking}
    \label{alg:sm3r}
    \small
    \KwIn{preference scores $\{\text{Pref}(u,i)\}_{i=1}^N$ of user $u$ on $N$ items, original item representations $\{\bm{c}_i\}_{i=1}^N$, disentangled aspects representations $\{\{\bm{v}_i^{a}\}_{a=1}^A\}_{i=1}^N$, trigger item $i_{t}$, trade-off weights$\{\lambda, \gamma\}$, output sequence length $L$\;}
    \KwOut{recommended item sequence $\mathcal{R} = \{i_1,i_2,\ldots, i_L \}$\;}
    Initialize: $l = 1$, accumulated relevance $ \bm{r}=\bm{0} $ \;
    Select first item $i_1$ by Eq.~\eqref{eq:one}\;

    \While{$l < L$}
    {
     \For{$a=1\rightarrow A$}
        {
       $ \bm{r}_a = \bm{r}_a + \Braket{\bm{v}_{i_l}^{a}, \bm{v}_{i_t}^{a}}$ ;  \tcp*[f]{accumulated relevance} \\
       }
      $l++$\;
      Update diversity weights $\bm{w}^D$ by Eq.~\eqref{eq:div_weight}\;
      Update relevance weights $\bm{w}^R$ by Eq.~\eqref{eq:rel_weight}\;
      \For{$i \in I\backslash \mathcal{R}_{1:l}$}
      {
        Calculate relevance scores $\text{Rel}(i, i_t)$ by Eq.~\eqref{eq:rel}\;
        Calculate diversity scores $\text{Div}(\mathcal{R}_{1:l-1}\cup i| i_t)$ by Eq.~\eqref{eq:div}\;
      }
      Select $i_l$ by Eq.~\eqref{eq:marginal_overall}\;
      
    }
    return $\mathcal{R}=\{i_1,i_2,\ldots, i_L \}$\
    
\end{algorithm}

\subsection{Discussion}\label{sec:discussion}

In this part, we make some discussions about the framework flexibility and complexity analysis.

\paratitle{Framework Flexibility}. 
Although we design the framework for relevant recommendation, it is easy to adapt it to other recommendation scenarios.
Actually, our framework presents a general way to balance multiple factors over the disentangled aspects. By zooming into our definition of relevance and diversity scores (see Eq.~\eqref{eq:rel} and Eq.~\eqref{eq:div}), we can see that it only involves the inner product operation between item embeddings.  Such a design is essentially model-agnostic, and any embedding-based recommendation models can be easily fit into our framework. 
As we do not introduce additional constraint on user/item embeddings, our framework can  be also implemented with different diversity algorithms. In this work, we adopt the classic MMR approach~\cite{carbonell1998use} as the backbone algorithm for diversity re-ranking, which can be replaced by Determinantal Point Process~(DPP)~\cite{chen2018fast} and 
Gram-Schmidt Process~(GSP)~\cite{huang2021sliding}. In our study, we test the performance of our framework with other diversity ranking methods (\eg Fast-DPP~\cite{chen2018fast} (a DPP variant) and SSD~\cite{huang2021sliding} (a GSP variant), but observe no significant improvement over MMR.  Due to the simplicity, we adopt MMR as the diversity ranking approach. 

\paratitle{Complexity Analysis.} The proposed self-balanced multi-aspect re-ranking algorithm iteratively selects items from candidates according to the similarity with selected items. Formally, it has comparable complexity with vanilla MMR~\cite{carbonell1998use} where the time complexity can be roughly estimated as $\mathcal{O} \big(L^2NAd\big)$ and space complexity is $\mathcal{O} \big(NAd\big)$ where $N$ is the number of candidate items and $L$ is the number of recommended items. The additional costs  are mainly attributed to the involving of the disentangled representations. To further reduce the costs, we can offline compute the embedding similarity (\eg using an efficient vector recall algorithm MIPS~\cite{shrivastava2014asymmetric}) and store the top similar embeddings via ID-based index. In this way, the time and space complexity can be reduced to $\mathcal{O} \big(LNAd\big)$ and $\mathcal{O} \big(NAd+LN\big)$ respectively. To sum up, our framework has comparable  complexity with those lightweight diversity re-ranking algorithms~(\eg Fast-DPP~\cite{chen2018fast} and SSD~\cite{huang2021sliding}) when  $M \ll L \ll N$.
\section{Experiments}
In this section, we first introduce the dataset which is collected from real relevant micro-video recommendation scenario in Section \ref{exp:dataset}. Then, the implementation details and experimental settings are presented in Section \ref{exp:setting} for reproducibility. Afterwards, the evaluation results of the proposed \ourmodel on both collected dataset and online $A/B$ test are shown in Section \ref{exp:offline} and \ref{exp:online} respectively. Finally, we conduct detailed analysis and case studies in Section \ref{exp:ablation} to further verify the effectiveness of proposed components.

\subsection{Datasets}\label{exp:dataset}
To the best of our knowledge, there is no available dataset customized for the relevant recommendation task. Therefore, we construct a new dataset, from KuaiShou app. We randomly sample 0.3 million active users and trace their logs in the \emph{Relevant Videos} page for 24 hours. We group the interaction data by user and trigger item to form different sessions, which consist of clicked items that both satisfy the user preferences and keep relevant to the trigger item. 
After filtering the items that are clicked fewer than 5 times, we collect nearly 4M interactions with 0.25M items and 0.6M sessions.
In order to capture feature-aware characteristics  of items, we collect the tags of each item as features. 
Besides, we initialize the embeddings of users and items with pre-trained parameters which are produced by the traditional collaborative filtering~(CF) model with large-scale training data to alleviate data sparsity problem. 
The detailed statistics of our collected dataset are shown in Table~\ref{tab:dataset}.

\begin{table}[]
\centering
\caption{Statistics of our dataset.}
\label{tab:dataset}
\vspace{-0.1in}
\begin{tabular}{@{}cccccc@{}}
\toprule
\#user & \#item & \#interaction & \#session & \#tag & $\frac{\#(\text{item},\text{tag})}{\#\text{item}}$ \\ \midrule
  369,329     &  241,143      &    4,016,182       &    648,777           & 97,840      &    9.1                \\ \bottomrule
\end{tabular}
\end{table}

\subsection{Offline Evaluation}\label{exp:offline}
To verify the effectiveness of the proposed \ourmodel in re-ranking, we first conduct offline experiments on the collected dataset. 
\subsubsection{Experimental Details}\label{exp:setting}
In the experiments, the number of embedding dimension is set to 128, which is the same as the hidden dimension of disentangled aspect representations. We employ a two-layer fully connected network with ReLU activation as our projection layer in DAE, where the hidden size is set to 128 and 64 respectively. The hyper-parameter, $\tau$, is set to 0.1 and we train the DAE module for 100 epochs with learning rate of 1e-4. The number of disentangled aspects is set to 5 and we conduct a grid search for all the trade-off weights in [0.001, 0.01, 0.1, 1], then report the best overall performance for all the compared methods. We use the dot product between the pre-trained user embeddings and item embedding as preference scores and relevance scores are computed from item embeddings. Then, the initial ranking is determined by the weighted sum of these two scores, and the weights are carefully tuned according to the recommendation performance.  
\subsubsection{Comparison Methods}
We compare our proposed framework with several baselines. The first one is  a Relevance-aware Ranking~(\textbf{RR}) algorithm~\cite{pang2017deeprank}, which jointly considers the user-item preference scores and item-item similarity scores for ranking. As we implement the diversity scoring function following \textbf{MMR}~\cite{carbonell1998use}, we select it as the main competitor which incorporates coarse-grained item-level diversity through re-ranking. Furthermore, we also select an efficient DPP-based algorithm, \textbf{FastDPP}~\cite{chen2018fast}, and a more recent diversified ranking algorithm, \textbf{SSD}~\cite{huang2021sliding}, as our compared methods. 
To explicitly verify the effectiveness of re-ranking and avoid sampling bias~\cite{krichene2020sampled}, we treat the whole item set as candidates for re-ranking~\cite{zhao2020revisiting}. 
Owing to the unaffordable time and space costs, we omit the comparison with other DPP-based algorithms~\cite{wilhelm2018practical,gartrell2017low}.

\subsubsection{Evaluation Metrics}
In addition to recommendation accuracy, our work also focuses on relevance and diversity. Specifically, we employ six metrics to comprehensively evaluate recommendation performance from different facets. Firstly, we exploit two commonly used metrics, \textbf{Recall} and \textbf{Mean Reciprocal Rank~(MRR)} to measure the accuracy. Then, we introduce a feature-aware relevance metric, \textbf{Mean Feature Hit Ratio~(MFHR)}, to represent the relevance between recommended items and the trigger item:
\begin{equation}
    MFHR=\frac{1}{|\mathcal{R}|}\sum_{i \in \mathcal{R}}\frac{|\mathcal{F}_i \cap \mathcal{F}_{i_t}|}{|\mathcal{F}_{i_t}|},
\end{equation}
where $\mathcal {R}$ denotes the recommended item set and $\mathcal{F}_i$ is the feature set of item $i$. Feature set of the trigger item is indicated as $\mathcal{F}_{i_t}$. This metric calculates the average hit ratio of features which can be seen as an indicator of relevance.
To evaluate the feature-aware diversity of recommended items, we modify two diversity metrics, \textbf{Feature Coverage Ratio~(FCR)} and \textbf{Intra List Average Distance~(ILAD)} to fit the task of relevant recommendation (similar metrics have been widely used by previous works~\cite{chen2018fast,wu2019pd}) as follows:
\begin{eqnarray}
    FCR =& \frac{|\bigcup\limits_{i\in \mathcal{R}} \mathcal{F}_i\cap \mathcal{F}_{i_t}|}{|\mathcal{F}_{i_t}|}, \\
    ILAD =& \frac{\sum_{i,j \in \mathcal{R}, i\neq j}1 - s(i, j)}{L(L-1)},
\end{eqnarray}
where $s(i,j) = \frac{|\mathcal{F}_i \cap \mathcal{F}_j \cap \mathcal{F}_{i_t}|}{|\mathcal{F}_i\cup \mathcal{F}_j \cap \mathcal{F}_{i_t}|}$ denotes the overlap ratio of features between item $i$ and $j$. Note that we only consider the overlapped features between recommended items and trigger item which shows the particularity of our task compared with conventional diversified recommendation task. Finally, we calculate the \textbf{F-score} over Recall and MFHR to find the balance between accuracy and relevance. For computational efficiency, we calculate metrics over the top-20 items after re-ranking. 

\subsubsection{Experimental Results}

\begin{table}[]
\centering
\caption{Results of offline experiments}
\label{tab:offline-exp}
\setlength{\tabcolsep}{1.1mm}{
\begin{tabular}{@{}c|cc|c|cc|c@{}}
\toprule
 \multirow{2}{*}{Methods}    & \multicolumn{2}{c|}{Accuracy} & Relevance & \multicolumn{2}{c|}{Diversity} & \multirow{2}{*}{F-score}  \\ \cline{2-6} 
     & Recall          & MRR         & MFHR      & FCR           & ILAD           &  \\ \midrule
RR   &  0.0894       & 0.0624     & 0.3511    & 0.9507        & 0.4222       & 0.1425   \\
FastDPP   & 0.0856        & 0.0601     &0.3497     & 0.9501     & 0.4269       & 0.1375    \\
SSD   & 0.0856        & 0.0586   & 0.3460     & 0.9493        & 0.4317     & 0.1373\\
MMR  &  0.1436      & 0.0818   & 0.3781    &  0.9968      &   0.4496     &   0.2082      \\
\ourmodel &  \textbf{0.1798} & \textbf{0.0968}   & \textbf{0.3936}   & \textbf{0.9989}       &  \textbf{0.4927}  & \textbf{0.2468}   \\\midrule
Improv. &  +25.2\% & +18.3\%   & +4.1\%   & +0.32\%       &  +9.6\%  & +18.5\%   \\
\bottomrule
\end{tabular}
}
\end{table}

The offline evaluation results of different methods on our industrial dataset are shown in Table~\ref{tab:offline-exp}. Based on the results, we can observe that:

(1) For those baselines that consider diversity, the performance of DPP and SSD is close to RR. A potential reason is they both assume that more orthogonality of the recommended items embeddings would result in larger diversity in the recommendation results. When the recommended items are highly relevant to the target item, the diversity score of these item embeddings will become relatively small compared to the relevance score. Therefore, it is difficult for those algorithms to effectively balance relevance and diversity in recommendation results.
In contrast, MMR outperforms the simple method RR in both accuracy and diversity, which indicates that using the dot product of embeddings as diversity score is more suitable than using orthogonality when item embeddings are similar in relevant recommendation. 
(2) By comparing our framework with all the baselines, it is clear to see that \ourmodel outperforms all baseline methods on all metrics. Different from baselines, our framework focuses on capturing fine-grained feature-aware relationships among items. As for the diversity metrics, the proposed \ourmodel performs better than baselines by a large margin on ILAD metric. These findings are consistent with our assumption that it is insufficient to only consider item-level diversity when candidate items are very similar in relevant recommendation. Finally, the F-score is also significantly improved, which further verifies that our framework can capture fine-grained preference characteristics of users for making personalized recommendations.

(3) Moreover, the improvement on both relevance and diversity indicates that our self-balanced multi-aspect re-ranker is able to adaptively adjust the weights of two targets in each aspect. These results also show that two proposed modules (\ie DAE and re-ranker) can coordinate well for improving the overall recommendation. 

\subsection{Further Analysis of \ourmodel}\label{exp:ablation}
In this section, we conduct more detailed analysis experiments to demonstrate the effectiveness of \ourmodel.

\subsubsection{Ablation Study}
To improve the performance of the relevant recommendation task, our proposed \ourmodel has incorporated several technical components. Next, we examine how each of them affects the final performance. Specifically, we consider the following variants of our framework for comparison through offline experiments: 
\begin{itemize}[leftmargin=*]
     \item  $\neg Disen$: the variant replaces the disentangled component with average pooling operation over feature embeddings, which produces a composite feature representation. In this variant, we only utilize this composite representation to promote feature-aware relevance and diversity.
     
     \item  $\neg SB$: the variant removes the self-balancing~(SB) strategy from the multi-aspect re-ranking algorithm.
     
   \item  $\neg Div$: the variant removes the fine-grained diversity target and only considers the relevance factor for re-ranking.
   
     \item  $\neg Rel$: this variant is opposite to the third case which ignores the relevance target and only maximizes the diversity score.
\end{itemize}

\begin{table}[]
\centering
\caption{Performance comparison of different variants.}
\label{tab:ablation}
\vspace{-0.1in}
\begin{tabular}{@{}l|ccc@{}}
\toprule
Variants  & Recall & MFHR & ILAD \\ \midrule
\ourmodel    & 0.1798       &  0.3936    & 0.4927     \\
\midrule
$\neg Disen$ & 0.1581      & 0.4035    & 0.4221    \\
$\neg SB $    & 0.1534       & 0.4209   &  0.3938    \\
$\neg Div$     & \textbf{0.1823}      & \textbf{0.4219}  & 0.4279     \\
$\neg Rel$     & 0.1388      &  0.3395   & \textbf{0.5365}    \\ \bottomrule
\end{tabular}
\end{table}
From the results in Table~\ref{tab:ablation}, we have the following observations. 
Firstly, replacing disentanglement aspect representations yields a large performance drop on both recall and diversity metrics. One possible reason is that simply introducing features into diversified re-ranking is unable to capture fine-grained feature interactions, thus resulting in sub-optimal performance.
Secondly, removing the self-balancing strategy would achieve a high relevance score but lead to a  decrease in diversity score. It indicates that  manual weights are difficult to effectively balance the diversity of multiple aspects, which further verifies the necessity of our proposed self-balancing strategy.
Finally, relevance and diversity are two vital factors that should be considered jointly. The two variants (\ie $\neg Div$ and $\neg Rel$) that only focus on one of the two factors achieve worse performance on the other factor. These results indicate that our proposed framework which jointly consider both relevance and diversity is effective to improve the overall performance. 
\begin{figure}[t]
	{
		\begin{minipage}[t]{0.45\linewidth}
			\centering
			\includegraphics[width=1\textwidth]{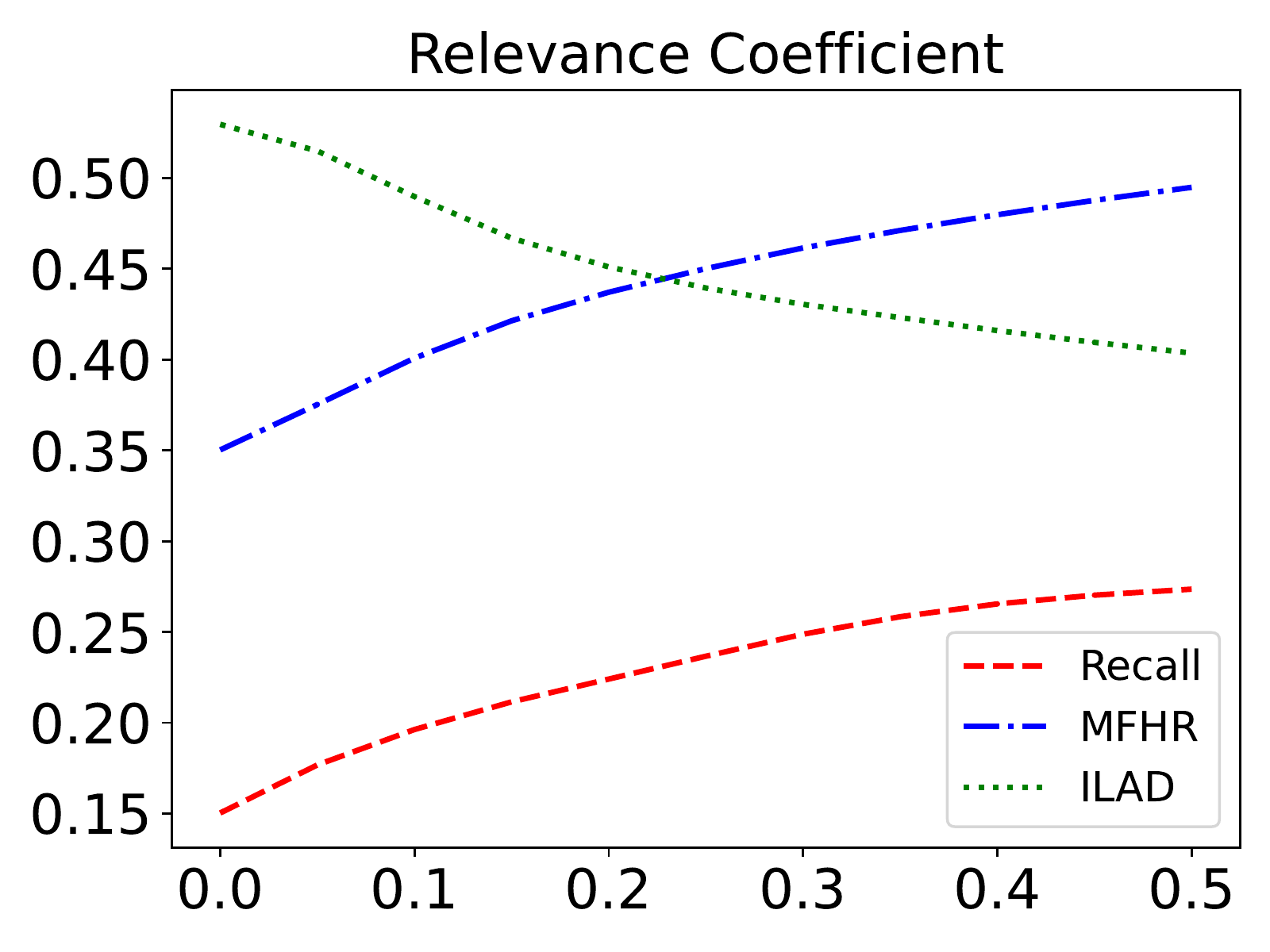}
		\end{minipage}
		\begin{minipage}[t]{0.45\linewidth}
			\centering
			\includegraphics[width=1\textwidth]{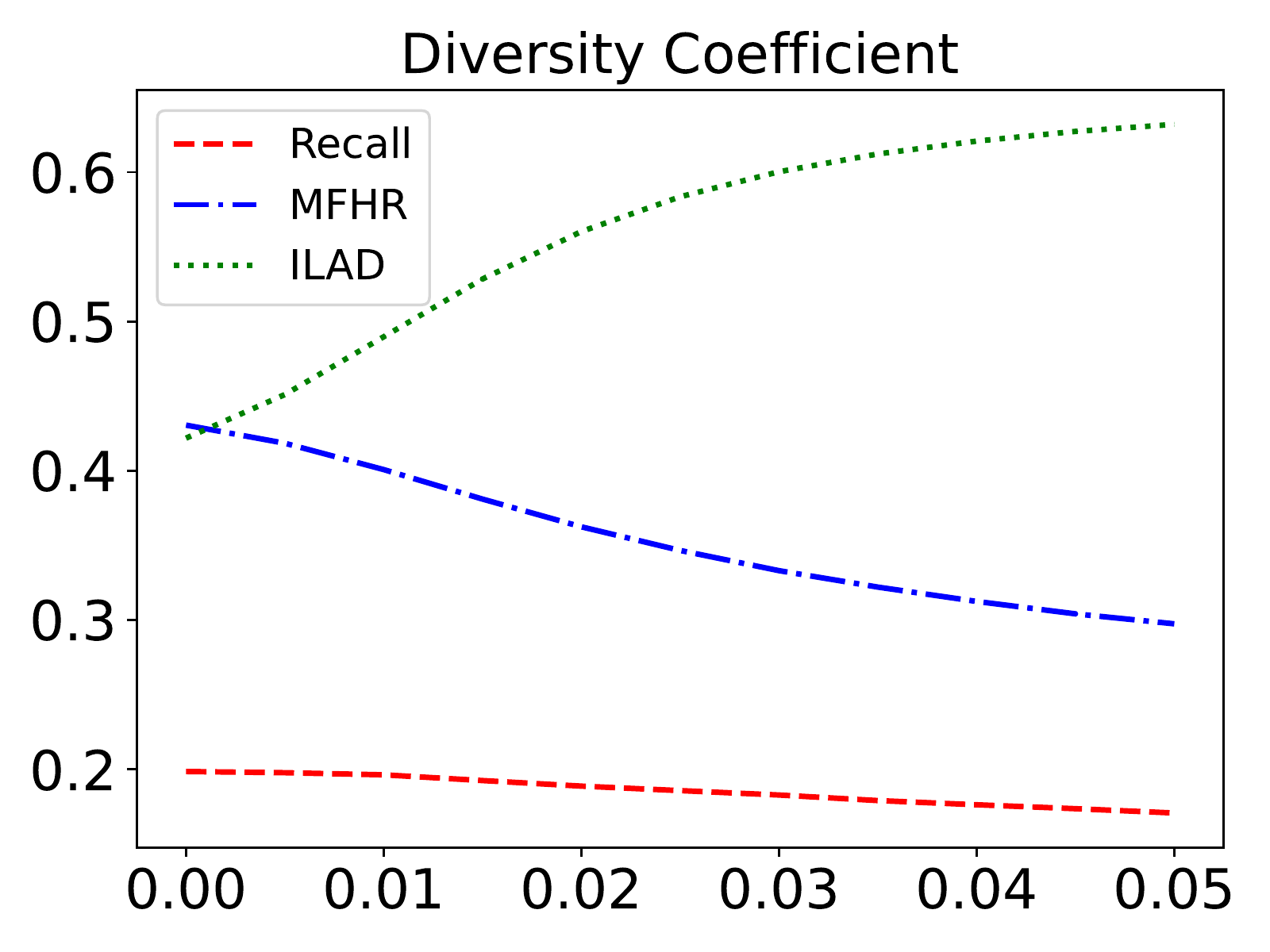}
		\end{minipage}
	}
	\caption{Performance on relevance and diversity w.r.t. different hyper-parameters. The left shows the results when varying $\lambda$ and the right shows the results when varying $\gamma$.} \label{fig:parameter}
\end{figure}

\subsubsection{Performance Tuning}
Since our framework utilizes two hyper-parameters $\lambda$ and $\gamma$~(parameters in Eq.~\eqref{eq:marginal_overall}) to balance the relevance and diversity, we continue to examine the effect of these two parameters. As we can see from Figure~\ref{fig:parameter},
as $\lambda$ becomes larger, MFHR and Recall increase, and ILAD decreases in the meantime. While, the increasing tuning of 
$\gamma$ lead to opposite results. This observation demonstrates the importance of the relevance factor in relevant recommendation which directly determines the final recommendation performance (\ie Recall). Moreover, compared to relevance, diversity is more sensitive to the choice of hyper-parameters. A possible explanation is that our proposed \ourmodel is better at depicting fine-grained diversity, so that it is more capable in obtaining high diversity with less loss  on accuracy. 

\subsubsection{Case Study}
\begin{figure}[t]
    \centering
    \includegraphics[width=0.45\textwidth]{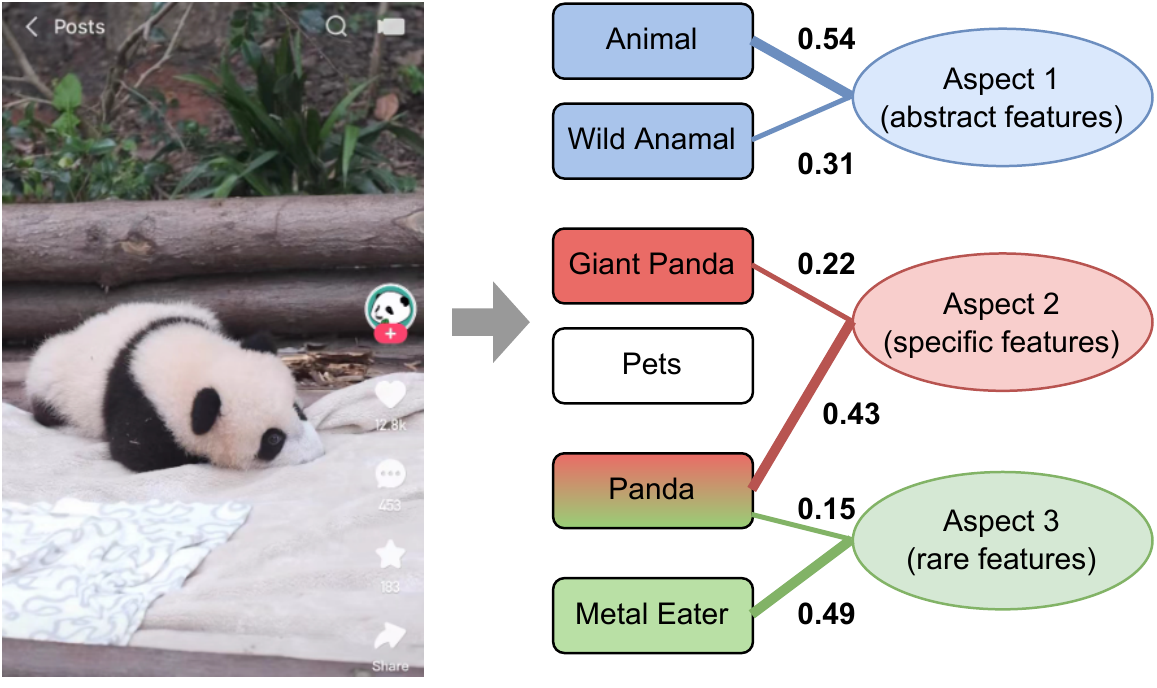}
    \caption{An example of the feature disentanglement. We plot six features and three disentangled aspects. Two primary features of each aspect are shown with attention weights.}\label{fig:case}
\end{figure}

In this part, we present one case for illustrating how our framework disentangles item features into different aspects. We randomly sample a video and visualize the attention weights of each aspect in Figure~\ref{fig:case}. As we can see, features are disentangled into different aspects according to their semantics. The first aspect focuses on abstract features like ``\emph{Animal}'' and ``\emph{Wild Animal}''. The second aspect captures fine-grained features like ``\emph{Panda}'' and ``\emph{Giant Panda}''. The third aspect extracts uncommon features where ``\emph{Metal Eater}'' is a nickname of panda. Note that the tag ``\emph{Pets}'' that is somehow redundant to this video has not been attended  by all aspects. This example qualitatively illustrates the effectiveness of our proposed DAE in learning disentangled feature representations.

\begin{table}[]
\centering
\caption{Results of online A/B tests on KuaiShou app.}
\label{tab:online}
\begin{tabular}{@{}c|ccc|c@{}}
\toprule
       & Watch Time & Video Play & Watched Tags &Very Relevant  \\ \midrule
\ourmodel    & +0.293\%  & +1.169\%   & +4.656\%  & +0.189\%    \\\bottomrule
\end{tabular}
\end{table}

\begin{figure}[t]
    \centering
    \includegraphics[width=0.45\textwidth]{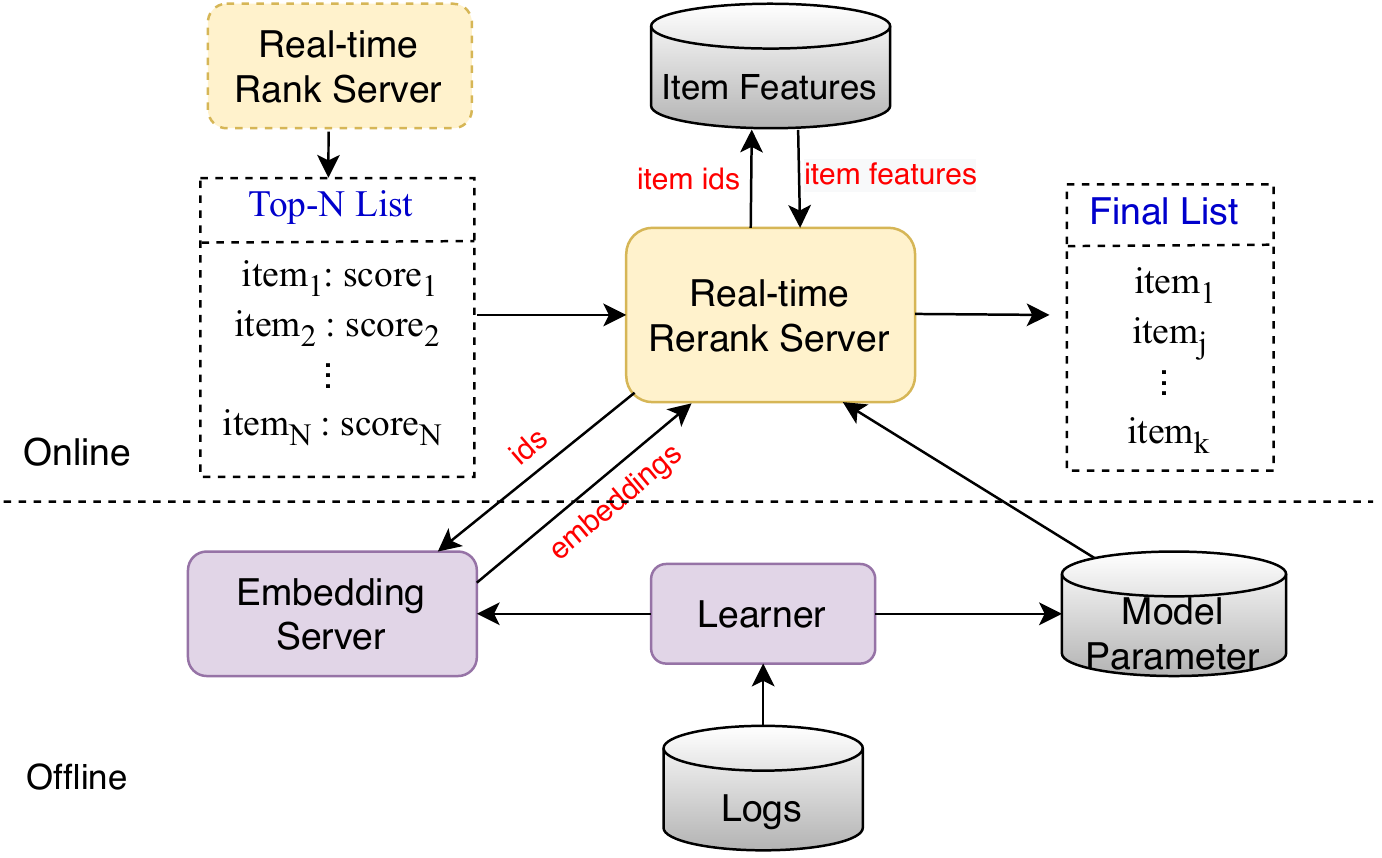}
    \caption{The workflow of our algorithm in relevant recommendation system.}
    \label{fig:rerank}
\end{figure}
\subsection{Online Deployment and $A/B$ Test}\label{exp:online}
To further verify the effectiveness of FDSB, we deploy it on the function of ``\emph{more relevant videos}'' on KuaiShou app for online $A/B$ test. The entire workflow is shown in Figure~\ref{fig:rerank}. 

We utilize the tags of videos as the item features, which are consistent with the offline experiments. To demonstrate the superiority of our framework in feature-aware relevance and diversity, we compare our framework with a baseline, MMR, to show the relative improvements. 

We evaluate the performance of the proposed framework in real application scenario of Kuaishou, with three metrics considering both user engagement and  recommendation diversity:  
(1) \emph{Watch Time}, (2) \emph{Video Play}, and (3) \emph{Watched Video Tags}. The first two metrics reflect the users' satisfaction, and the third metric is  commonly used for diversity. Furthermore, to examine the relevance between recommended items and trigger item, we randomly deliver a questionnaire to a small population of users during the recommendation service. The users are required to rating on relevance according to a 3-point Likert scale~\footnote{\url{https://en.wikipedia.org/wiki/Likert_scale}}, \ie the fourth metric. We report the proportion that users assign a 3-point rating, which means ``Very Relevant''.
The $A/B$ test is conducted for seven consecutive days and we report the average performance on these four metrics.  

From Table~\ref{tab:online}, we can have the following observations. Firstly, \ourmodel achieves significant improvements in both \emph{Watch Time} and \emph{Video Play}, indicating that our framework is able to increase user loyalty (more per-video watching time and more video plays, \ie watching videos). Secondly, the number of tags for watched videos also largely increases, which suggests we promote the diversity in the recommendation results. Thirdly, more users respond with ``\emph{Very Relevant}'' in our questionnaires. This result shows that \ourmodel can recommend more relevant videos and further explains why \ourmodel has longer \emph{Watch Time} in the relevant recommendation. Overall, by incorporating item features and self-balanced strategy, both the diversity in recommendation results and the relevance between recommended items and trigger item are enhanced, showing that our framework has made a good trade-off between relevance and diversity in the relevant recommendation task. 

\section{Related work}
In this section, we review related works on relevant recommendation, aspect-aware recommendation and diversified recommendation.

\paratitle{Relevant Recommendation}.
To the best of our knowledge, there are few works that focus on the relevant recommendation~\cite{xie2021real}. In~\cite{xie2021real}, the authors proposed a framework named R3S to jointly rank relevant items and decide whether to display the items. They designed a multi-critic multi-gate mixture-of-experts strategy to jointly model the information of user, trigger and context. 
While this method can not effectively distinguish features of different aspects, which is important to  promote the diversity for relevant recommendation.  Moreover, this work has not explicitly model the  recommendation
 diversity. Besides, R3S is an end-to-end model that requires more  labeled data to optimize, while our \ourmodel is a lightweight re-ranking algorithm which is easy to optimize and deploy. Furthermore, it is flexible to integrate our method with existing methods such as R3S for relevant recommendation. For example, our approach can take the predictions of R3S as the preference scores for making more accurate recommendations.

\paratitle{Aspect-aware Recommendation}. 
Several studies have been proposed to extract aspects from side information to guide the recommendation~\cite{guan2019attentive,cheng2018aspect}. Most of existing works design aspect-aware architecture to understand text (\eg reviews and item description), which is able to model users and items from different aspects for fine-grained matching. For example, \citet{guan2019attentive} proposed a well-designed aspect-level attention module to select related description from user and item reviews.
As a recently proposed technique, disentangled representation learning~\cite{tran2017disentangled,yang2021safe} has been widely applied to extract latent semantic dimensions from data in a variety of fields, \eg recommender systems~\cite{zheng2021disentangling,fan2021lighter, wu2021fairrec}.
Most of the previous works learn disentangled representations with graph-based model~\cite{chen2021decomposed,zheng2021disentangling,mu2021knowledge} (either based on interaction graph or knowledge graph) or self-attention model~\cite{fan2021lighter}. 
 Furthermore, to alleviate the data bias in recommendation, an additive causal model was proposed to learn disentangled representations of interest and conformity~\cite{zheng2021disentangling}. In the task of sequential recommendation, LightSANs~\cite{fan2021lighter} introduced decomposed self-attention and decoupled position encoding for context-aware representations.
Different from existing works, we consider learning feature-level aspects for modeling relevance and diversity in a specific setting of relevant recommendation. 

\paratitle{Diversified Recommendation}.
Recently, the diversity of recommendations has received much attention in the research community~\cite{wu2019recent}, which is proved to be  
 a NP-hard problem~\cite{toth2000optimization}.
Several works conducted the diversified recommendation as an optimal selection problem and utilized   post-processing  ~\cite{carbonell1998use,chen2018fast,antikacioglu2017post} or adversarial sampling~\cite{sun2020framework,zheng2021dgcn,ye2021dynamic} for enhancing the diversity. 
Overall, there are two mainstream categories of diversified recommendation, namely aggregation diversity~\cite{ge2010beyond,zhang2021model} and individual diversity~\cite{cheng2017learning,sha2016framework}. 
In this work, we focus on the individual diversity of relevant recommendation. 
Recently, determinantal point process~(DPP)~\cite{gillenwater2014expectation} has been widely used as a principled approach to conducting diversified recommendation. 
To tackle the problem of high complexity in these methods, a sampling algorithm was proposed based on eigen-decomposition~\cite{gillenwater2014approximate} for approximation. Furthermore, SSD~\cite{huang2021sliding} leveraged the Gram-Schmidt process to reduce the space complexity. 
Actually, diversity was considered in the field of information retrieval at a much earlier time, and a classical work MMR ~\cite{carbonell1998use} proposed to greedily select docs based on the similarity with both the query and the currently selected docs, inspiring a number of extension works~\cite{qin2013promoting,sha2016framework}. 
Different from the above approaches, our work is the first to consider both diversity and relevance in re-ranking stage for relevant recommendation. Although we currently adopt the MMR approach for modeling the diversity, FDSB is a model-agnostic framework and can be instantiated with other diversity methods~\cite{chen2018fast, huang2021sliding}.  

\section{Conclusion and Future work}
In this work, we study the task of relevant recommendation from a real application scenario, where relevance and diversity are two important factors that need to be considered by recommenders. In order to better characterize item-to-item relations and capture feature-aware diversity, we propose a re-ranking framework, \textbf{FDSB}, based on a multi-factor ranking approach. There are two modules, namely disentangled attention encoder~(DAE) and self-balanced multi-aspect re-ranker in FDSB. In the DAE, we learn disentangled representations from rich item features. Then we adaptively balance the relevance and diversity with a multi-aspect re-ranker. Both offline evaluation and online  $A/B$ test have demonstrated the effectiveness of the proposed framework.

For future work, we will study how to capture both relevance and diversity in an end-to-end ranking model for relevant recommendation. Besides, we will also consider modeling feature-aware fine-grained diversity in other recommendation tasks or scenarios.

\begin{acks}
This work was partially supported by National Natural Science Foundation of China under Grant No. 61872369,
Beijing Natural Science Foundation under Grant No. 4222027, and Beijing Outstanding Young Scientist Program under Grant No. BJJWZYJH012019100020098.
This work was also partially supported by Beijing Academy of Artificial Intelligence~(BAAI).
Xin Zhao is the corresponding author.
\end{acks}


\bibliographystyle{ACM-Reference-Format}
\bibliography{ref}

\end{document}